\documentclass[structabstract]{aa}  
\usepackage{graphicx}
\usepackage{txfonts}
\usepackage{natbib}
\usepackage{multirow}

\begin{document}
   \title{Evolution of microflares associated with bright points in coronal holes and in quiet regions}


   \author{S. Kamio \and W. Curdt \and L. Teriaca \and D. E. Innes}
\institute{Max-Planck-Institut f\"ur Sonnensystemforschung (MPS),
Max-Planck-Str. 2, 37191 Katlenburg-Lindau, Germany\\
\email{skamio@spd.aas.org}
}

   \date{Received ; accepted }

 
  \abstract
   {}
   {We aim to find similarities and differences
between microflares at coronal bright points found in quiet regions and coronal holes, and to study their relationship with large scale flares.}
   {Coronal bright points in quiet regions and in coronal holes were observed with {\it Hinode}/EIS using the same sequence.
Microflares associated with bright points are identified from the X-ray lightcurve.
The temporal variation of physical properties was traced in the course of microflares.}
   {The lightcurves of microflares indicated an impulsive peak at hot emission followed by an enhancement at cool emission, which is compatible with the cooling model of flare loops.
The density was found to increase at the rise of the impulsive peak,
supporting chromospheric evaporation models.
A notable difference is found in the surroundings of microflares;
diffuse coronal jets are produced above microflares in coronal holes
while coronal dimmings are formed in quiet regions.}
   {The microflares associated with bright points share common characteristics to active region flares.
The difference in the surroundings of microflares are caused by open and closed configurations of the pre-existing magnetic field.}

   \keywords{Sun: corona --- Sun: coronal mass ejections (CMEs) --- Sun: UV radiation}

   \maketitle
%

\section{Introduction}

\begin{figure*}
 \centering
 \includegraphics[width=12cm]{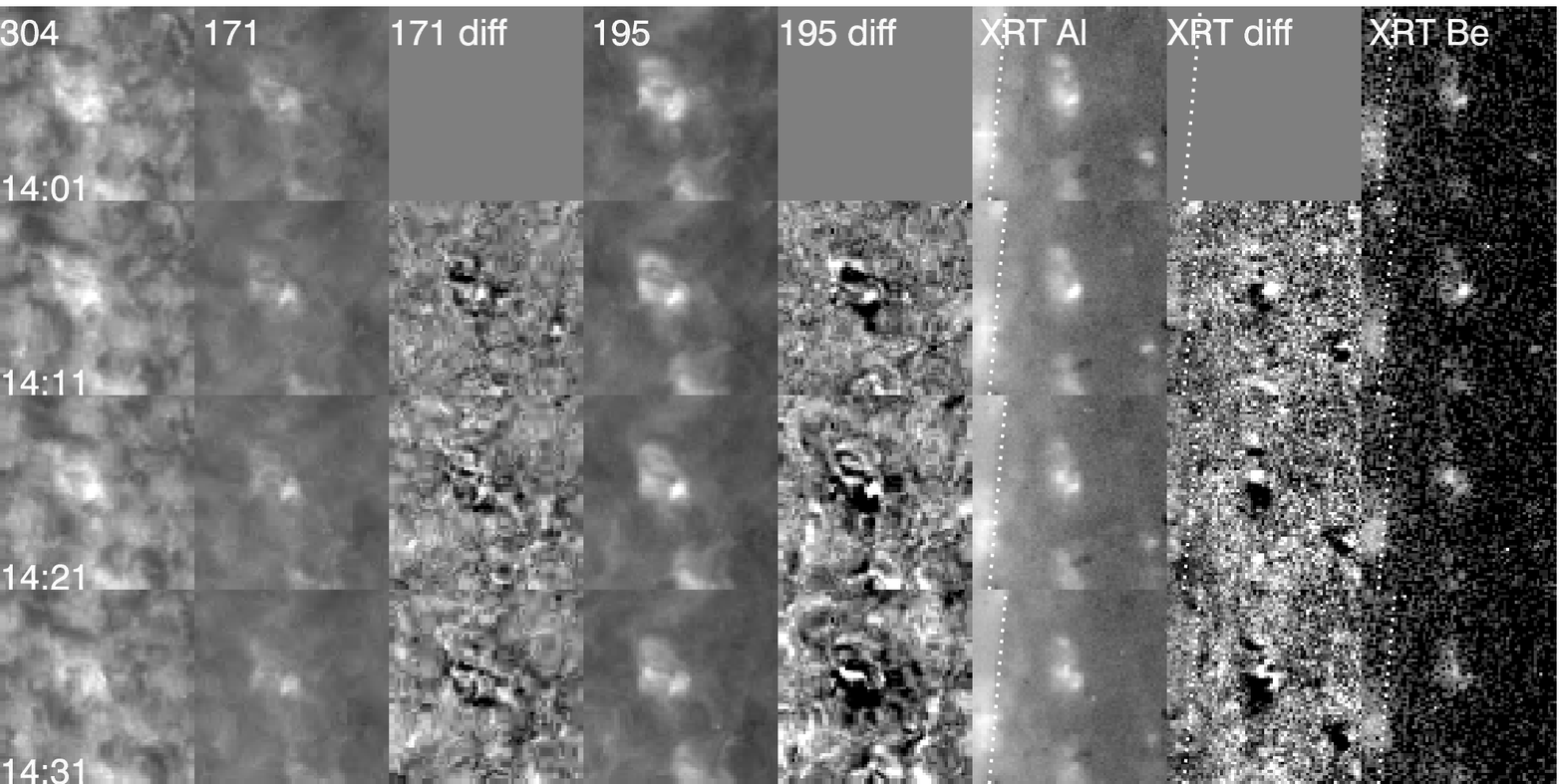}\\
 \vspace*{0.5cm}
 \includegraphics[width=12cm]{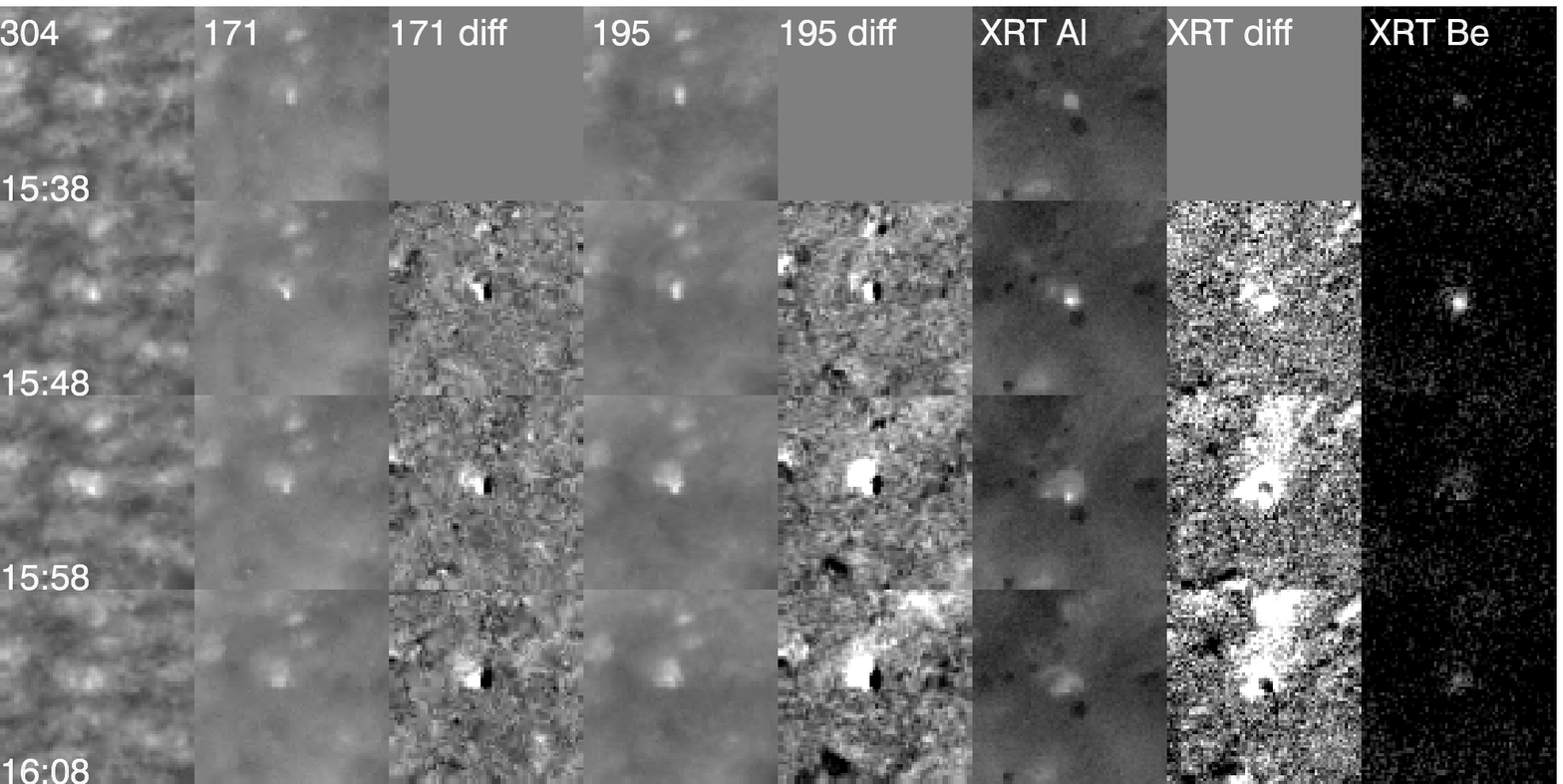}
 \caption{{\it Top:} Temporal evolution of a bright point in
a quiet region (QR1).
Columns display time series of SECCHI/EUVI images in
$\lambda$30.4~nm (\ion{He}{ii}), $\lambda$17.1~nm (\ion{Fe}{ix/x}), and
$\lambda$19.5~nm (\ion{Fe}{xii}) and XRT images with
Al\_poly and Be\_thin filters with 10 min intervals.
Difference images are indicated by 'diff' on columns.
Dimension of the each panel is 100\arcsec$\times$100\arcsec.
Dotted lines on the XRT images mark the solar limb.
{\it Bottom:} Temporal evolution of a bright point in a coronal hole (CH1).}
 \label{fig:image}
\end{figure*}

Coronal bright points are small roundish or loop-shaped features
seen in X-ray and UV emissions all over the Sun.
Their typical size is 5--10\arcsec, and they last for several
hours or longer \citep{golub1974}.
Thanks to the high sensitivity of the X-ray telescope
(XRT; \citealt{golub2007}) on {\it Hinode} \citep{kosugi2007},
bright points are found to be highly dynamic and time varying
\citep{cirtain2007, kotoku2007}.
It is known that bright points occasionally
undergo a sudden increase in emissions
on a time scale of several minutes \citep{strong1992, koutchmy1997}.
In this paper, we call these transient brightenings as in the corona
microflares, and focus on the microflares associated with bright points
in quiet regions and coronal holes.

The study of microflares is important
for understanding the process of energy release in the corona,
since they ubiquitously occur in quiet regions and coronal holes.
Although the size of microflares is much smaller than
that of large flares in active regions,
they share some common characteristics.
Bright points are formed above small magnetic concentrations in the photosphere
\citep{golub1977}, while active regions are a manifestation of
large scale magnetic flux.
Recent observations have revealed small scale
ejections which resemble coronal mass ejections (CMEs) from active regions
\citep{innes2009,innes2010}.
It is worthwhile to study the evolution of microflares
and examine if their properties agree with the standard flare model
\citep{carmichael1964,sturrock1966, hirayama1974, kopp1976}.
It is essential to trace the physical properties
such as temperature and density in the course of microflares in
order to compare microflares with the flare model.

A sequence of XRT images show that many jets originate from microflaring
bright points in coronal holes, while
coronal jets are not common to bright points in quiet regions.
One question to be answered is whether these two types of microflares
are caused by the same mechanism or not.
The distinct difference
gives insights into interactions between
microflares and magnetic fields in the surroundings.
Jets in coronal holes have been extensively studied
\citep{savcheva2007, shimojo2007, subramanian2010}.
They are clearly observed against the dark background
in coronal holes.
However, a detailed comparison of microflares
in quiet regions and coronal holes has not been made yet.

In this respect, we carried out observations to trace the temporal evolution
of microflares at bright points.
Determining physical properties of microflares is of great importance.
To compare bright points in a quiet region and in a coronal hole,
both targets were monitored with the same observing sequence
near the solar limb.
The paper is organized as follows;
the methodology is explained in Sect. 2;
the observations and data reduction procedures are described in Sect. 3;
the properties of microflares in a quiet region and in a coronal hole are
presented in Sect. 4.
the formation of microflares and their similarity to active region flares
are discussed in Sect. 5;
and the conclusions are summarized in Sect. 6.


\section{Methodology}

\begin{figure}
 \centering
 \includegraphics[width=8cm]{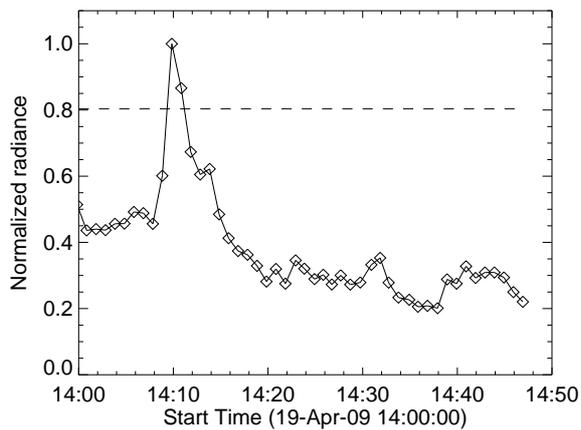}
 \caption{An X-ray lightcurve of a bright point indicating
a microflare at 14:10 UTC.
  Dashed line marks 3$\sigma$ level of the light curve.}
 \label{fig:lx00}
\end{figure}

The primary purpose of our study is to find similarities and differences
in the evolution of microflares in quiet regions and in coronal holes.
To this end, the same observing sequence was carried out at
four locations near the limb:
quiet regions in the east and in the west and
polar coronal holes at the north and at the south.

We use coronal imagers to trace the 
evolution of microflares and coronal structures
in the surroundings.
A sequence of XRT images was employed to identify
microflares at coronal bright points.
Since the X-ray images were recorded at 1 min cadence, they had the best
temporal coverage of the evolution.
An X-ray lightcurve of a bright point is created by
spatially integrating X-ray flux over the bright point.
We selected impulsive enhancements
exceeding the 3$\sigma$ level of the lightcurve,
where $\sigma$ is the standard deviation.
Figure \ref{fig:lx00} shows an X-ray lightcurve of a bright point.
A peak at 14:10 UTC exceeding 3$\sigma$ level is selected as a microflare
(QR1 in Table \ref{table:bp}),
and the peak time is defined as the time of the microflare.
A 4\arcsec$\times$2\arcsec\, summing was applied to the XRT data
in order to match the EIS resolution.
The brightest pixel in the summed XRT data at the time of the microflare is selected for
following analysis as it represents the microflare.

In addition, time series of EUV images obtained by
SECCHI/EUVI \citep{howard2008,wuelser2004}
on {\it STEREO} \citep{kaiser2008} are
analyzed to study the coronal structures associated with
these microflares.
Difference images are created to detect small changes in emission,
which is a widely used technique since the discovery
of coronal dimmings \citep{thompson1998}.

At the same time, physical properties of microflares
are deduced.
The EUV Imaging Spectrometer \citep[EIS;][]{culhane2007}
on {\it Hinode}
carried out repeated fast raster scans to study the evolution
of microflares at different temperatures
ranging from the transition region to the corona.
Although the temporal resolution of the data is not
good as compared to XRT, the spectra allow us to deduce the temperature
evolution of the microflares.
The observed time difference between the hot and cool emission lines
is compared with a simple loop cooling model by \citet{cargill1995},
by analogy with large scale flares.
In addition, the line ratio technique was applied to detect density
enhancements in the corona,
which we expected from chromospheric evaporation
in the standard flare model.
Our method is to identify a distinct microflare and to track the
temporal evolution, hence the number of events detected in our study
is not sufficient for a statistical study.

\section{Observations and data reduction}

Four sets of coordinated observations were carried out
on 19 and 21 April 2009.
No active regions appeared at that time.
The observing sequence is composed of
rapid raster scans by EIS and
time series of coronal images for
three hours in each location.

\begin{table}
\caption{List of emission lines observed with EIS}             
\label{table:linelist}      
\centering                          
\begin{tabular}{l c c}        
\hline\hline                 
Ion & $\lambda$ / nm & $\log T_e$ / K \\    
\hline                        
\ion{Fe}{viii} & 18.660 & 5.6 \\
\ion{Fe}{xii}  & 18.688 & 6.1 \\
\ion{Fe}{xi}   & 19.281 & 6.1 \\
\ion{O}{v}     & 19.291 & 5.4 \\
\ion{Fe}{xii}  & 19.512 & 6.1 \\
\ion{Fe}{xvii} & 25.487 & 6.6 \\
\ion{He}{ii}   & 25.632 & 4.7 \\
\ion{Si}{x}    & 26.106 & 6.1 \\
\ion{Fe}{xv}   & 28.416 & 6.3 \\

\hline                                   
\end{tabular}
\end{table}

EIS on {\it Hinode} repeated sparse rasters at 6 min cadence.
Spectra of emission lines listed in Table \ref{table:linelist}
were recorded with the 2\arcsec~ wide slit and at exposure
time of 25 s.
Emission lines were selected to cover a wide range of temperature
within a limited telemetry volume.
The resultant pixel resolution is 2\arcsec$\times$2\arcsec\, 
after summing along the slit to increase the signal
to noise ratio.
Each raster scan consists of 12 exposures with 4\arcsec\, step size,
and an area of 44\arcsec $\times$ 320\arcsec.
The duration of the sequence was one hour
due to the eclipse periods during the {\it Hinode} Sun-synchronous orbit.

EIS data are calibrated with the standard procedures
provided in the Solar Software tree \citep[SSW;][]{freeland1998}.
A single Gaussian fit was applied to spectra
to deduce radiance, Doppler velocity, and line width.
A single Gaussian results in a reasonable fit in
quiet conditions.
Spectral profiles taken during the microflares occasionally
deviate from Gaussian profiles because of enhancements
in the wings, which are presented in Sect. 4.3.
As for EIS, the Doppler shift is deduced after compensating for
the instrumental effect caused by temperature variations
\citep{kamio2010b} and
using the rest wavelengths of emission lines identified by
\citet{brown2008}.
The contribution of the \ion{Si}{x} $\lambda$25.637~nm blending in
the \ion{He}{ii} $\lambda$25.632~nm is subtracted by using
the other \ion{Si}{x} $\lambda$26.106~nm emission
\citep{kamio2009}.
Although a coronal blending remains
in the red wing of the spectrum, \ion{He}{ii} is the
dominant emission at that wavelength.

Time series of coronal images were obtained to study
the evolution of the microflares.
XRT recorded soft X-ray images with Al\_poly and Be\_thin filters.
The former is sensitive to low temperature coronal emission
down to $1\times10^6$ K and shows coronal structures
in quiet regions.
The latter detects the higher temperature corona and
normally results in a weak signal in quiet regions.
It was intended to capture the hot component in coronal
bright points.
A pair of X-ray filter images were obtained every 60 s.
The area of the XRT images were 384\arcsec$\times$384\arcsec,
and the EIS area was covered.
Co-alignment between EIS and XRT was performed by comparing the
\ion{Fe}{xii}~$\lambda$19.512~nm radiance map from EIS and
XRT Al\_poly images.
We estimate a co-alignment error of 4\arcsec,
which corresponds to the scan step of the EIS data.
Other emission lines are registered
after compensating for the north -- south offset of
the EIS spectra \citep{kamio2010b}.

{\it STEREO} EUVI obtained
images with filters of $\lambda$17.1~nm (\ion{Fe}{ix/x}),
$\lambda$19.5~nm (\ion{Fe}{xii}), and
$\lambda$30.4~nm (\ion{He}{ii}).
EUVI images with the set of filters were recorded at
a cadence of 10 min.
A coordinate conversion is necessary to find the bright
points observed with XRT in the SECCHI/EUVI images.
The separation angles between the Earth and
{\it STEREO}-A and B spacecrafts
were $46.7^{\circ}$ and $47.0^{\circ}$, respectively.

Since the {\it Hinode} spacecraft is in low-Earth orbit \citep{kosugi2007},
XRT images are regarded as Earth view.
Assuming that a coronal bright point was located on the
solar surface,
Heliocentric Earth Ecliptic (HEE)\footnote{
http://secchi.nrl.navy.mil/wiki/uploads/Main/coordinates.pdf}
coordinates of the bright point are deduced from its apparent
position in XRT images.
The coordinates are transformed into {\it STEREO} spacecraft views
by using the {\it STEREO} SPICE software package provided in the SSW tree.
The position of the coronal bright point on the {\it STEREO} EUVI
images is determined by taking $x$ and $y$ coordinates in the {\it STEREO}
spacecraft view.

\section{Results}

In our data set, 10 microflares are
identified in X-ray images (7 in quiet region and 3 in coronal hole)
and are listed in Table \ref{table:bp}.
In the following, two representative events are described:
one in a quiet region and the other in coronal hole.
They are labelled as QR1 and CH1 in Table \ref{table:bp}.
Similarities and differences between the quiet region and the polar coronal
hole and the implications for the coronal structures are discussed in Sect. 5.

\subsection{Coronal structures}

The top panel of Fig. \ref{fig:image} shows a time series of a bright point
in a quiet region near the East limb (QR1 in Table \ref{table:bp}).
Each column presents a 100\arcsec$\times$100\arcsec\,
section of SECCHI/EUVI images in
$\lambda$30.4~nm (\ion{He}{ii}), $\lambda$17.1~nm (\ion{Fe}{ix/x}), and
$\lambda$19.5~nm (\ion{Fe}{xii}), and XRT images with
Al\_poly and Be\_thin filters.
As the Al\_poly images show, the bright point was located just 20\arcsec~
inside the limb.
The top row shows pre-event images of the bright point at 14:01 UTC.
Enhanced emission was detected at the center of all panels at 14:11 UTC,
which suggests that the microflare took place in
a wide temperature range.
A significant emission increase is detected with the XRT Be\_thin filter,
which is rare in a quiet region.
This is indicative of a hot component produced by the microflare,
since the Be\_thin filter is only sensitive to the high temperature corona.

Difference images in $\lambda$17.1~nm (\ion{Fe}{ix/x}), and
$\lambda$19.5~nm (\ion{Fe}{xii}), and XRT Al\_poly are
derived by subtracting pre-event images at 14:01 UTC
to emphasize the temporal variation.
Emissions of the microflare reached their peaks just before 14:11 UTC
while a significant dimming was observed in the
surrounding.
A careful analysis of the high cadence (1 min) XRT time series shows that
the dimming started at 14:08 UTC, when the X-ray flux of the microflare
underwent a sharp rise.
The dimming continued to expand until 14:31 UTC.
As the {\it STEREO}-B spacecraft was separated from the Earth
by $47.0^{\circ}$, SECCHI/EUVI provided a better coverage of the region
close to the limb.
The difference images in $\lambda$19.5~nm show that
the size of the dimming expanded to 30\arcsec~.
The dimming was also seen in $\lambda$17.1~nm, but was
restricted to 10\arcsec, from the bright point.
In Table \ref{table:bp}, the first columns present heliocentric coordinates
of the bright points, which were mostly observed close to the limb.
An apparent diameter of microflare is determined
by using original XRT images without pixel summing. This is because
the resolution of a summed pixel is not sufficient for
measuring apparent size.
A typical apparent diameter of the microflare is 3 -- 5$\times10^8$ cm.
The apparent areas of the microflares correspond to
the areas covered by pixels in the XRT Al\_poly images brighter
than the 3$\sigma$ level.

The bottom panel of Fig. \ref{fig:image} shows a time series of a bright point
in a coronal hole near the North limb (CH1).
Increased emission is detected at the center of all images at 15:48 UTC,
which is similar to the quiet region event, QR1.
A major difference is that no coronal dimming is found in the surroundings.
Instead, a diffuse jet is seen above the microflare
in the XRT Al\_poly difference image.
The diffuse jet extended to 50\arcsec at 15:58 UTC.
The difference images indicate darkenings only in the bright point,
indicating a small emission decrease from the pre-flare structure.

\subsection{Lightcurves}

\begin{figure}
 \centering
 \includegraphics[width=6cm]{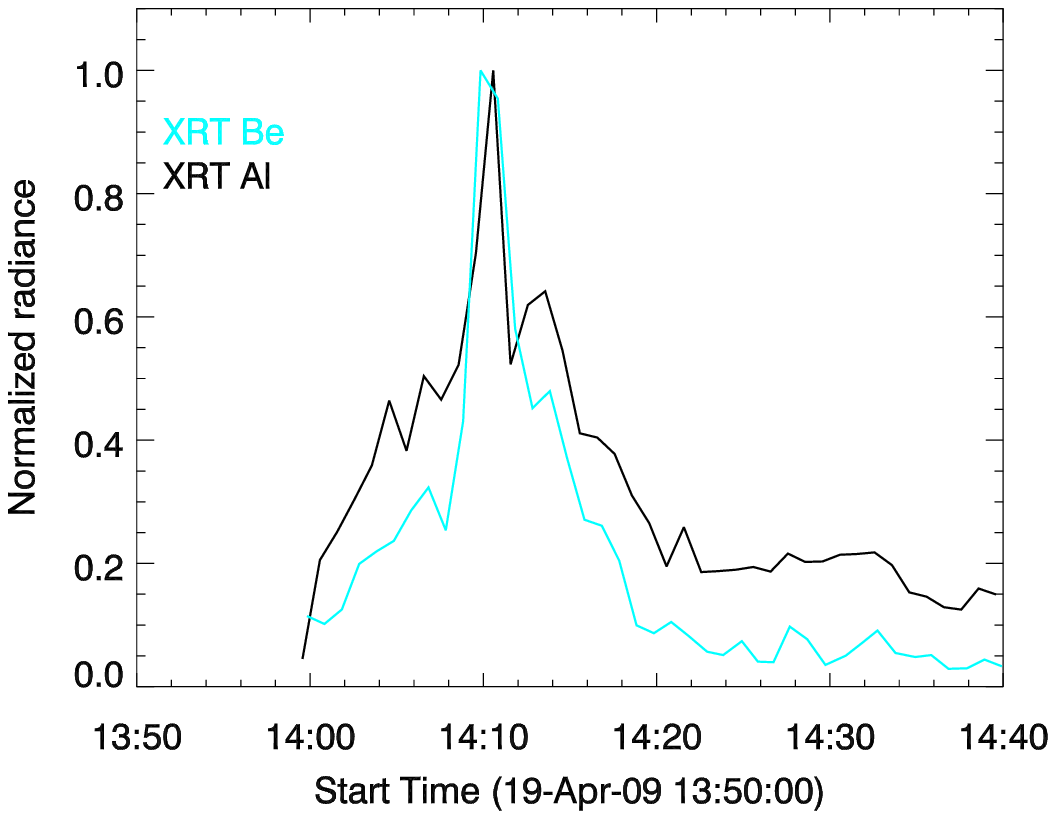}
 \includegraphics[width=6cm]{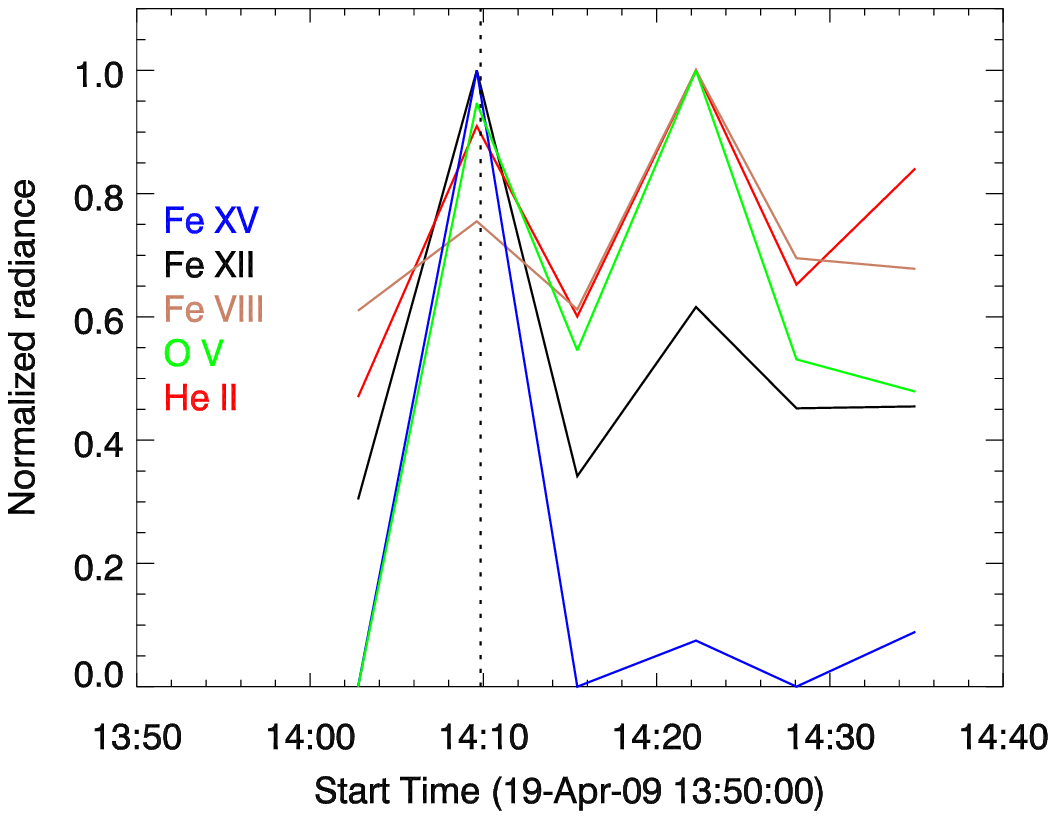}
 \includegraphics[width=6cm]{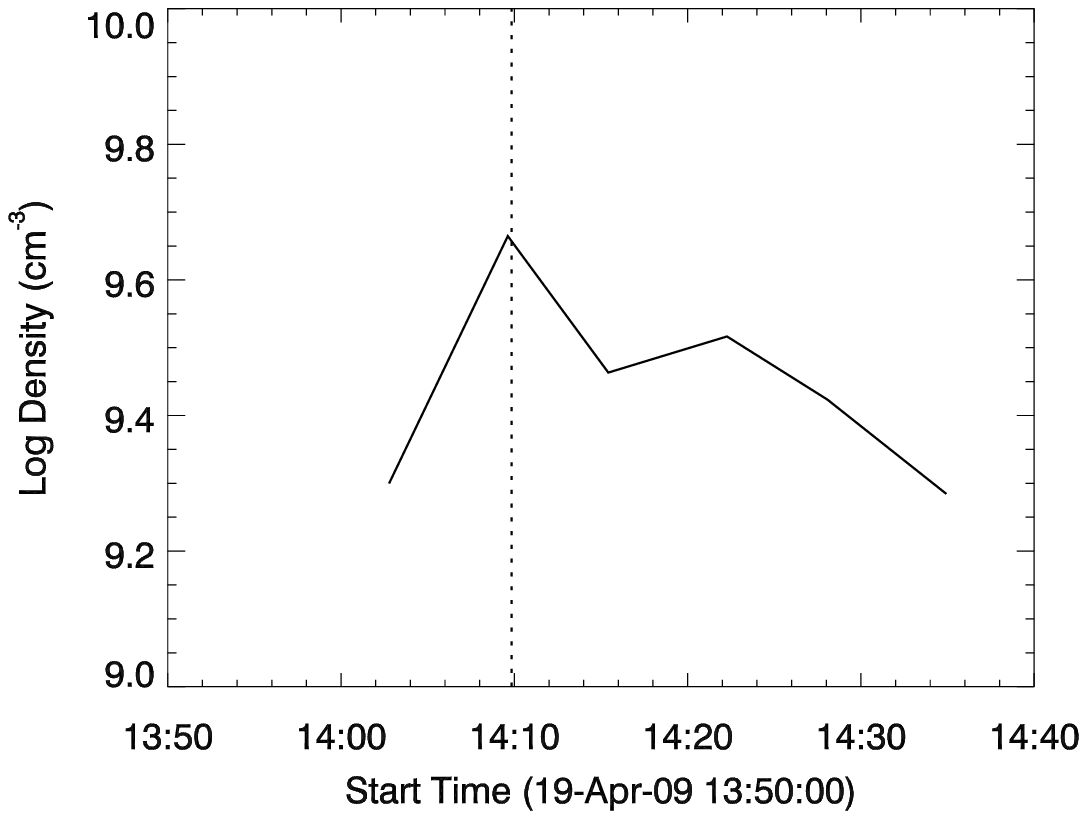}
 \caption{Normalized lightcurves of the microflare in a quiet region (QR1).
  {\it Top:} X-ray lightcurves from XRT Be\_thin and Al\_poly filters.
  {\it Middle:} lightcurves from EIS emission lines.
  {\it Bottom:} the electron density (cm$^{-3}$) deduced from the
   emission ratio of \ion{Fe}{xii} $\lambda$18.688~nm / $\lambda$19.512~nm.}
 \label{fig:lc01}
\end{figure}

\begin{figure}
 \centering
 \includegraphics[width=6cm]{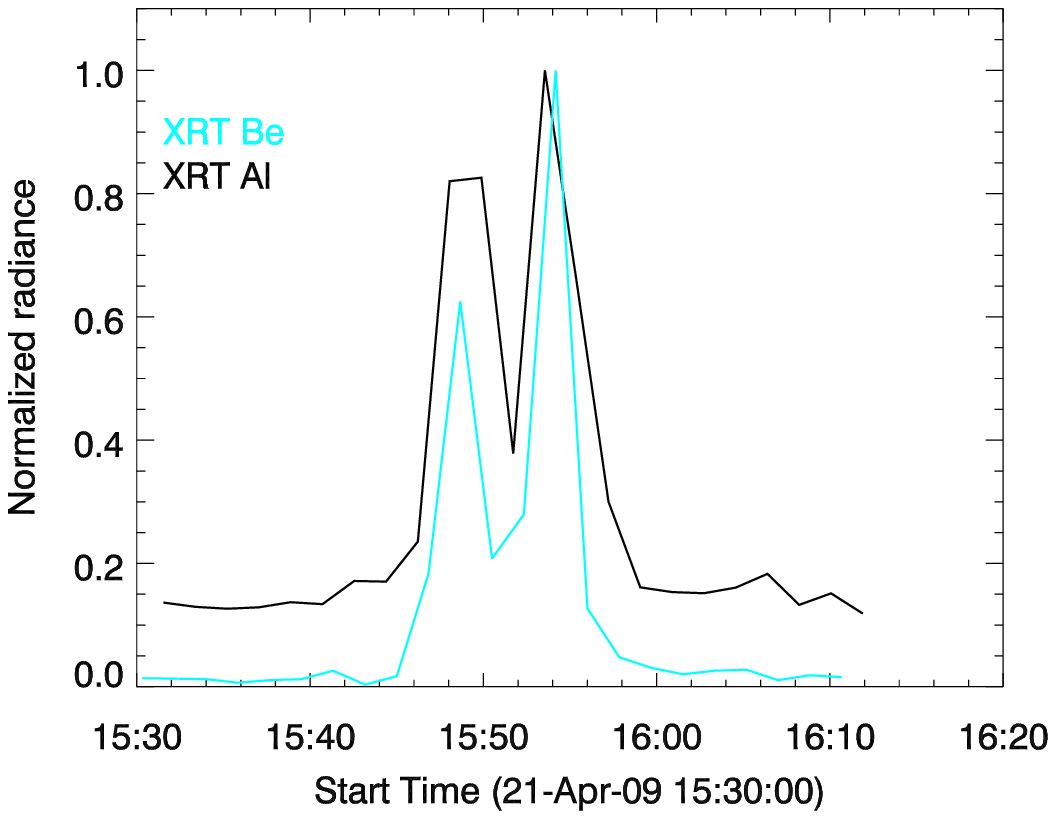}
 \includegraphics[width=6cm]{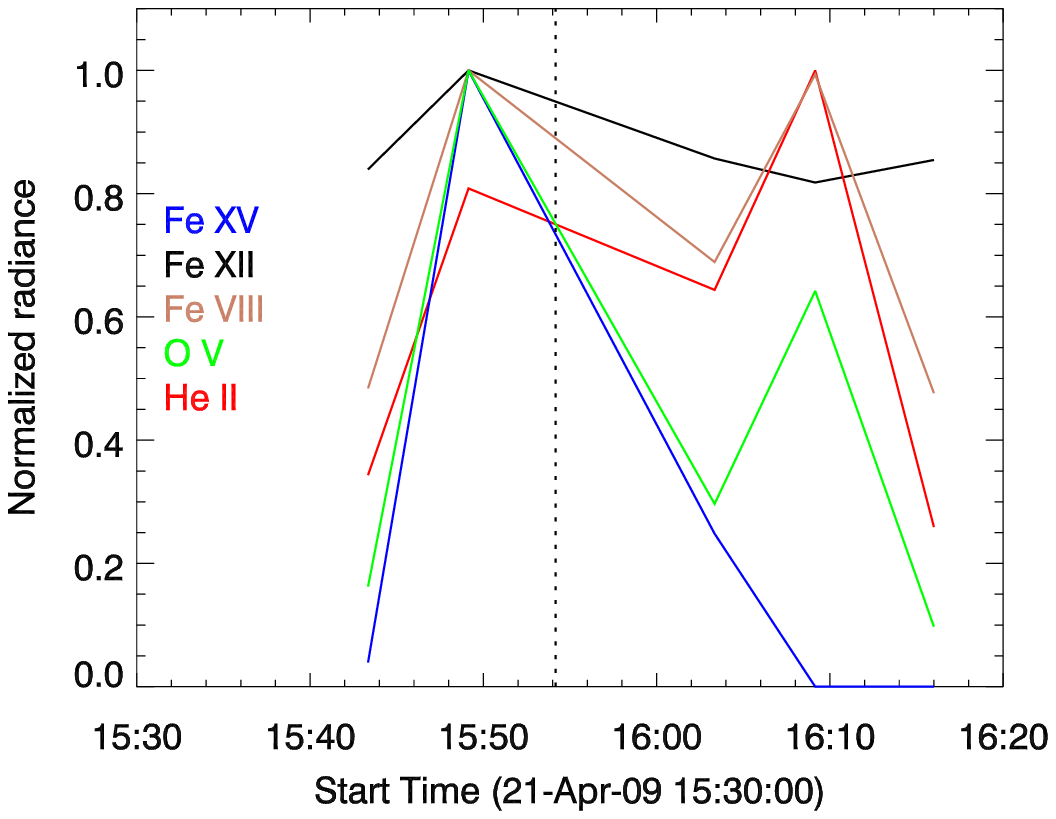}
 \includegraphics[width=6cm]{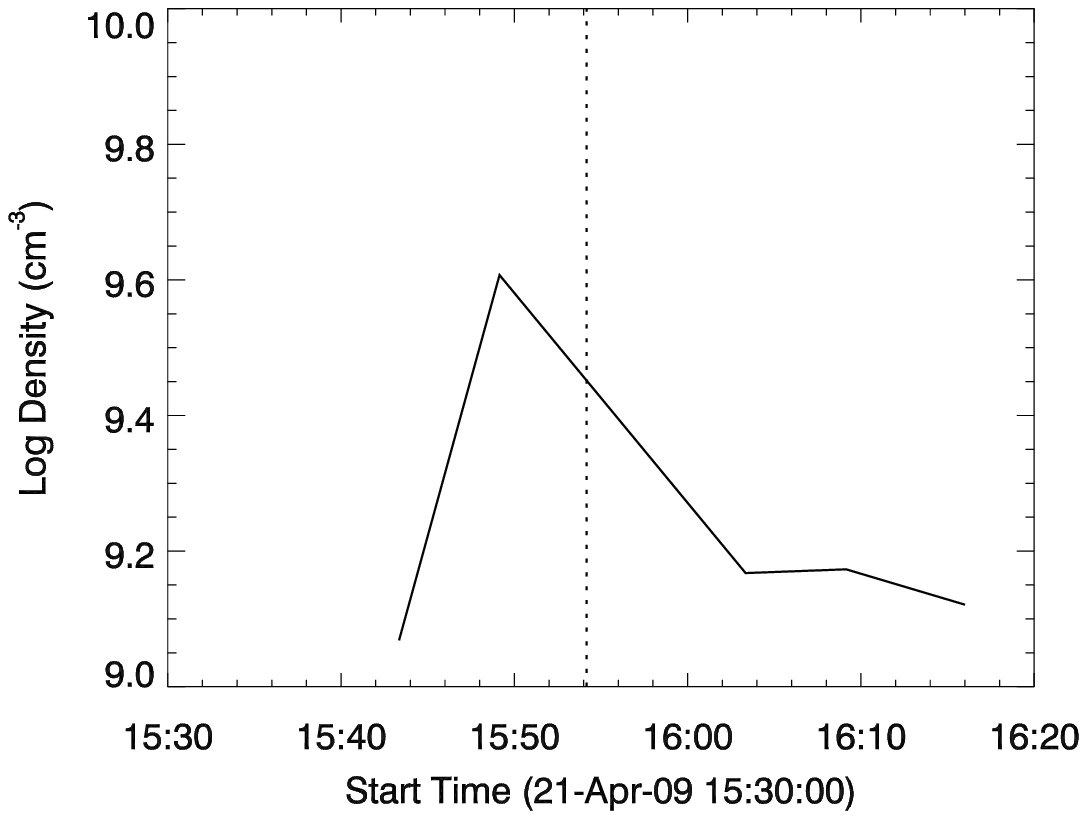}
 \caption{The same as Fig. \ref{fig:lc01} but for the microflare in a coronal hole (CH1).}
 \label{fig:lc08}
\end{figure}

\begin{figure*}
 \centering
 \includegraphics[width=16cm]{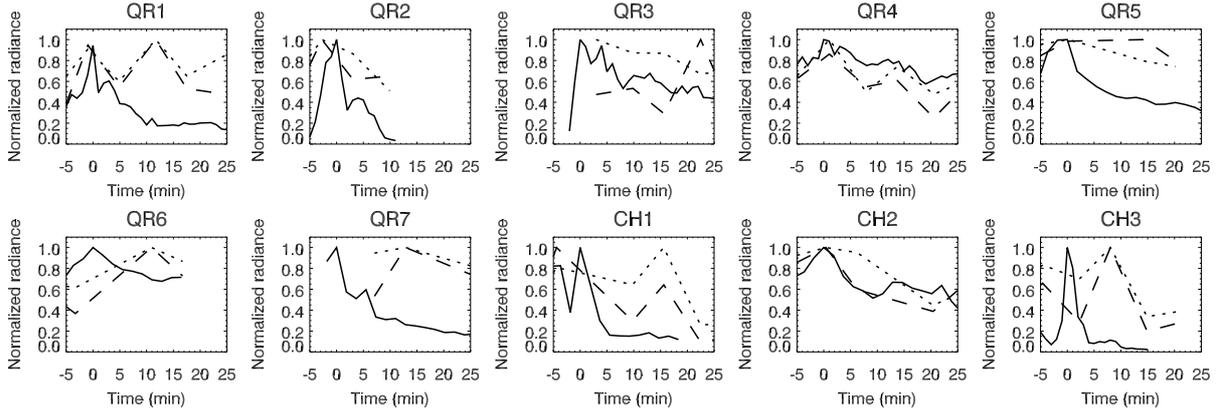}
 \caption{Normalized lightcurves of XRT Al\_poly (solid line),
\ion{He}{ii} (dotted line), and \ion{O}{v} (dashed line) for all microflares.
Horizontal axis indicates the time relative to the XRT Al\_poly peak.
}
 \label{fig:lc_all}
\end{figure*}

Figure \ref{fig:lc01} presents lightcurves of the bright point QR1.
The X-ray lightcurves are deduced from 4\arcsec$\times$2\arcsec~
area, which match the resolution of the EIS data.
Shortly after the {\it Hinode} observations start,
a rapid rise in emission the main peak started at 14:08 UTC.
The maximum flux is attained in the Be\_thin and Al\_poly
filters at 14:10 UTC.
Fluxes decrease to their pre-event level by 14:25 UTC,
hence the duration of the enhanced radiance was 25 min.

Co-spatial lightcurves of emission lines observed by
EIS are also plotted in Fig. \ref{fig:lc01}.
Note that the temporal resolution of EIS is 6 min.
 
An enhanced emission was observed in all emission lines
at 14:10 UTC, almost in coincidence with the X-ray peak.
A sharp rise in \ion{Fe}{xv} reflects the fact that
the \ion{Fe}{xv} emission is normally absent
in quiet regions and is significant only during the event.
The peak is also observed in other emission lines,
though its enhancement is less pronounced than the \ion{Fe}{xv}.
This multi-temperature peak is regarded as a heating
from the chromosphere to the corona
during the microflare.
After the X-ray peak, another emission peak was detected in
\ion{He}{ii} and \ion{O}{v} at 14:22 UTC, while 
the \ion{Fe}{xv} and \ion{Fe}{xii} emission remained
at pre-event levels.
The delayed peak in cool emission is interpreted as
cooling of the hot plasma produced by the microflare.
This hypothesis is examined in Sect. 4.7.
The delay time is defined as the time between the X-ray peak
and the following peak in cool emission lines, either
\ion{O}{v} or \ion{He}{ii}.
For the QR1, the delay time was 12 min.
The hottest recorded emission was in the \ion{Fe}{xv} line,
and no emission was detected in \ion{Fe}{xvii}.
This can used to set the upper limit to the temperature of the microflare
($T_e < 4\times10^6$K ).

Figure \ref{fig:lc08} shows the lightcurves for the bright point
in the coronal hole (CH1).
The X-ray fluxes indicate two subpeaks;
the first subpeak at 15:49 UTC and the second is at 15:53 UTC.
An EIS exposure coincided with the first subpeak, which
underwent enhanced emissions in all emission lines.
The following exposure missed the second peak
and captured its decay phase.
As the \ion{Fe}{xii} emission gradually decreased,
delayed enhancements were found in
the \ion{O}{v} and \ion{He}{ii} lines.
The delay between the largest X-ray peak and
the cool emissions was 15 min.

It must be noted that the accuracy of the delay times can only be determined
to within the EIS temporal resolution of 6 min.
Nevertheless, it is worthwhile to check if the delayed cool
emissions are common phenomena.
Figure \ref{fig:lc_all} presents lightcurves of all events
for selected emissions.
Delay times are measured for other events and summarized in
Table \ref{table:bp}.
Delayed cool emission were found for 7 out of the 10 microflare events.
No delay time is determined for QR2, QR4, and CH2, since
no clear increase in emission was observed in the cool lines.

\subsection{Spectra of microflares}

\begin{figure}
 \centering
 \includegraphics[width=8cm]{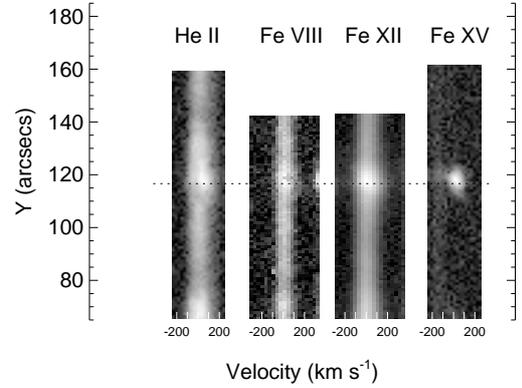}
 \caption{Spectra in \ion{He}{ii} $\lambda$25.632~nm,
  \ion{Fe}{viii} $\lambda$18.660~nm,
  \ion{Fe}{xii} $\lambda$19.512~nm, and
  \ion{Fe}{xv} $\lambda$28.416~nm recorded across
   the microflare QR1 at its peak emission time.
  The horizontal axis shows Doppler velocity relative to the line center position.
  Dashed line marks the location of the microflare.
  }
 \label{fig:sp01}
\end{figure}

\begin{figure}
 \centering
 \includegraphics[width=8cm]{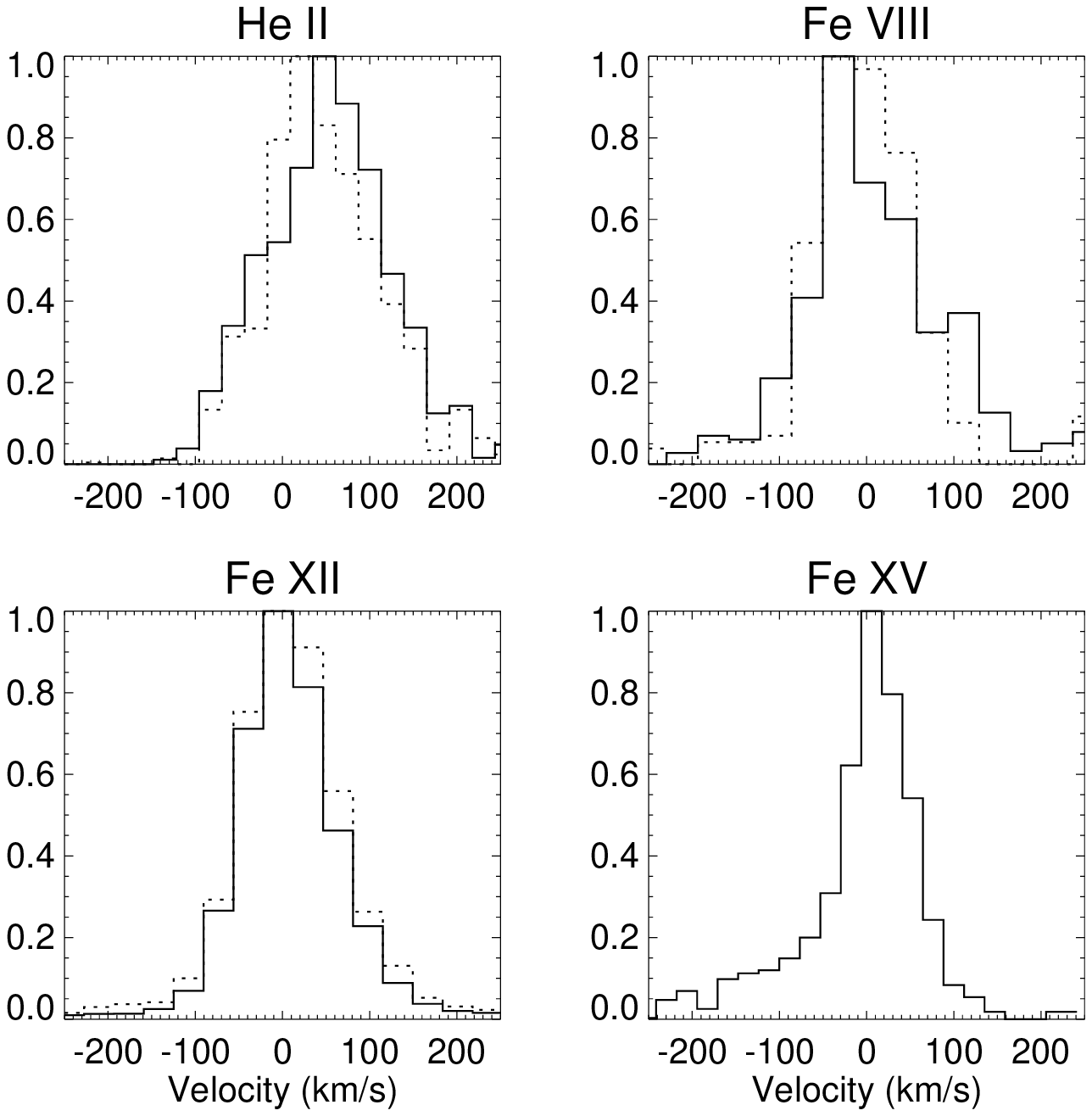}
 \caption{Normalized spectral profiles at the microflare QR1
  marked by dashed line in Fig. \ref{fig:sp01}.
  The profiles are normalized by the peak emission.
  The dotted lines indicate profiles in
  a quiet region neighboring the microflare.}
 \label{fig:pro01}
\end{figure}

Figure \ref{fig:sp01} presents spectra obtained by EIS
coinciding with the peak of the microflare labelled QR1 at 14:10 UTC.
Normalized profiles are plotted in Fig. \ref{fig:pro01}.
The dotted line show a spectral profile in
a quiet region 10\arcsec apart from the microflare, except for \ion{Fe}{xv},
for which no \ion{Fe}{xv} emission was detected in the surrounding.
The \ion{He}{ii} line indicated a redshift of 40 km~s$^{-1}$.
The \ion{Fe}{viii} profile at the microflare was broader than
in the quiet region, indicating that turbulent motions exist in the microflare.
The \ion{Fe}{xii} emission increased at the microflare, and
a blueshift of about 10 km~s$^{-1}$ was found.
The \ion{Fe}{xv} spectrum shows its emission only at the microflare.
It suggests that the hot component is confined to the microflare
as the emission is absent in the surroundings.
The \ion{Fe}{xv} spectrum shows a noticeable enhancement of the
blue wing up to $-200$ km~s$^{-1}$.
The blue wing enhancements are attributed to a horizontal flow rather
than a radial flow, as the radial component has a smaller contribution
to the line of sight velocity
due to the location of the bright point on the limb.

\begin{figure}
 \centering
 \includegraphics[width=8cm]{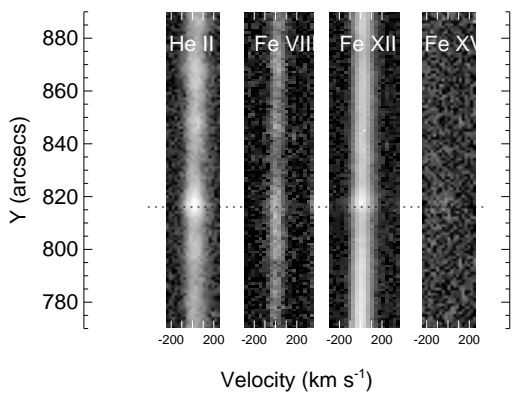}
 \caption{The same as Fig. \ref{fig:sp01} but for the microflare CH1.}
 \label{fig:sp08}
\end{figure}

\begin{figure}
 \centering
 \includegraphics[width=8cm]{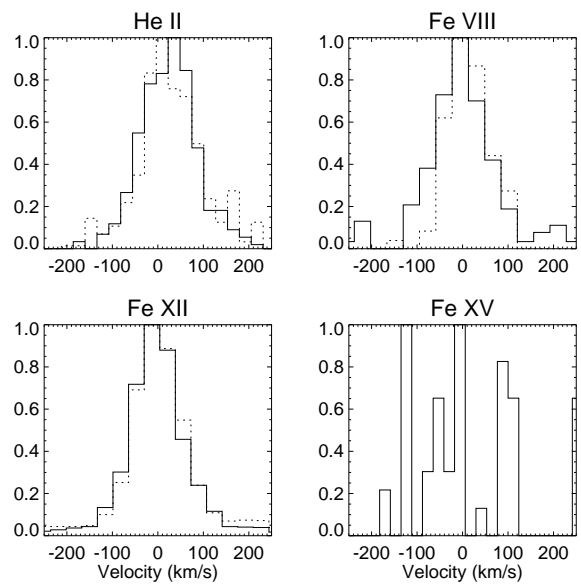}
 \caption{Normalized spectral profiles of the microflare CH1
  indicated with dashed line in Fig. \ref{fig:sp08}.}
 \label{fig:pro08}
\end{figure}

Spectra from the microflare CH1 are displayed in
Figs. \ref{fig:sp08} and \ref{fig:pro08}.
A broadening is found in the \ion{Fe}{viii} line.
The extended wings suggest that turbulent motions exist in the microflare.
\ion{Fe}{xv} exhibits a noisy pattern because of the low
count rate. However, it amounts to a significant
emission after integrating over the wavelength range
(Fig. \ref{fig:lc08}).

Generally the emission in coronal holes is weaker than in quiet region.
Extended wings of the profiles, which suggest turbulent motions in microflares,
are observed in both quiet region and coronal hole.

\subsection{Electron densities}

The electron density $n_e$ in the microflare is deduced
from the emission ratio of
\ion{Fe}{xii} $\lambda$18.688~nm and $\lambda$19.512~nm,
and $n_e$ is derived from
the lightcurves of the two emission lines, which are processed in the same
way as in Sect. 4.2.
The line ratio analysis is used because the $n_e$ can be
deduced without any assumptions about the volume or filling factor of the plasma.
The CHIANTI atomic database version 6.01 \citep{dere2009b, dere1997}
estimates that the ratio of the \ion{Fe}{xii} emission lines varies from
$1.1\times10^{-2}$ to $1.7$
with an electron density range of
$10^8$ -- $10^{12}$ cm$^{-3}$.
Electron densities derived from the \ion{Fe}{xii} ratio are
plotted in the bottom panels of Figs. \ref{fig:lc01} and \ref{fig:lc08}.

The density of QR1 before the microflare was
$2\times10^9$~cm$^{-3}$ (Fig. \ref{fig:lc01}).
It reached $4\times10^9$~cm$^{-3}$ at the
time of the peak emission in \ion{Fe}{xii}.
The uncertainty of the density estimated from the radiance measurement
error is less than 20\%,
hence the density increase is significant.
The density gradually decreased to the pre-event value by 14:37 UTC
when the enhanced emission diminished.
Before the microflare CH1, the density was
$1\times10^9$~cm$^{-3}$,
and attained $4\times10^9$~cm$^{-3}$ at the time of
peak emission (Fig. \ref{fig:lc08}).
The density dropped back to $1\times10^9$~cm$^{-3}$ when the following exposure
was taken.
\citet{dere2008} reported a typical density of
$4\times10^9$~cm$^{-3}$ in bright points, a value which falls within
the QR1 density range,
however, the temporal variation of the density was not studied by \citet{dere2008}.
Our study shows density enhancements associated with microflares.
Both QR1 and CH1 indicate that the density increases
reaching a peak which coincides with the peak of the \ion{Fe}{xii} emission.
The values of $n_e$ derived at the \ion{Fe}{xii} peak emission of the other
microflares identified are presented in Table \ref{table:bp}.

\citet{shimizu1995} carried out statistical study of microflares in
an active region and derived a typical density of
$2\times10^9$~cm$^{-3}$ -- $2\times10^{10}$~cm$^{-3}$.
Although our dataset is not sufficient for a statistical study,
derived densities are below $5\times10^{9}$~cm$^{-3}$,
hence the densities of microflares in quiet region and coronal holes
tend to be lower than those in active regions.

\subsection{Emission Measure of microflares}

%
\begin{table*}
\caption{Properties of microflares}             
\label{table:bp}      
\centering          
\begin{tabular}{c c c c c c c c c c c}     
\hline\hline       
\multicolumn{2}{c}{Location} & \multicolumn{2}{c}{Coordinates}  &  diameter & area & density & delay & cooling time & total EM & radiative loss rate\\ 
& & X & Y & (cm) & (cm$^2$) & (cm$^{-3}$) & (min) & (min) & (cm$^{-3}$) & (erg s$^{-1}$) \\
\hline
 \multirow{7}{*}{QR}
 & 1 & $-925$\arcsec & $115$\arcsec & $4\times10^8$ & $1.1\times10^{17}$ & $4\times10^9$ & 12 & 9 & $1.3\times10^{44}$ & $3.9\times10^{22}$ \\ 
 & 2 & $-955$\arcsec & $-10$\arcsec & $3\times 10^8$ & $8.4\times10^{16}$ & $5\times10^9$ & - & 8 & $2.7\times10^{43}$ & $9.2\times10^{21}$ \\  
 & 3 & $-940$\arcsec & $-30$\arcsec & $5\times10^8$ & $1.8\times10^{17}$ & $2\times10^9$ & 22 & 13 & $1.4\times10^{44}$ & $4.7\times10^{22}$ \\  
 & 4 & $-940$\arcsec & $-146$\arcsec & $5\times10^8$ & $2.2\times10^{17}$ & $2\times10^9$ & - & 14 & $1.4\times10^{44}$ & $5.5\times10^{22}$ \\  
 & 5 & $932$\arcsec & $20$\arcsec & $4\times10^8$ & $1.6\times10^{17}$ & $2\times10^9$ & 15 & 12 & $3.9\times10^{43}$ & $1.6\times10^{22}$ \\  
 & 6 & $930$\arcsec & $-60$\arcsec & $5\times10^8$ & $1.6\times10^{17}$ & $1\times10^9$ & 11 & 13 & $8.1\times10^{43}$ & $2.8\times10^{22}$ \\
 & 7 & $10$\arcsec & $735$\arcsec & $5\times10^8$ & $1.7\times10^{17}$ & $4\times10^9$ & 13 & 11 & $5.3\times10^{43}$ & $1.8\times10^{22}$ \\
 \hline
\multirow{3}{*}{CH}
 & 1 & $-12$\arcsec & $820$\arcsec & $4\times10^8$ & $9.2\times10^{16}$ & $4\times10^9$ & 15 & 8 & $2.3\times10^{43}$ & $7.6\times10^{21}$ \\
 & 2 & $10$\arcsec & $910$\arcsec & $5\times10^8$ & $2.3\times10^{17}$ & $2\times10^9$ & - & 15 & $1.2\times10^{44}$ & $4.4\times10^{22}$ \\
 & 3 & $0$\arcsec & $-940$\arcsec & $3\times10^8$ & $9.7\times10^{16}$ & $3\times10^9$ & 8 & 9 & $4.9\times10^{42}$ & $1.9\times10^{21}$ \\
\hline                  
\end{tabular}
\end{table*}

\begin{figure}
 \centering
 \includegraphics[width=8cm]{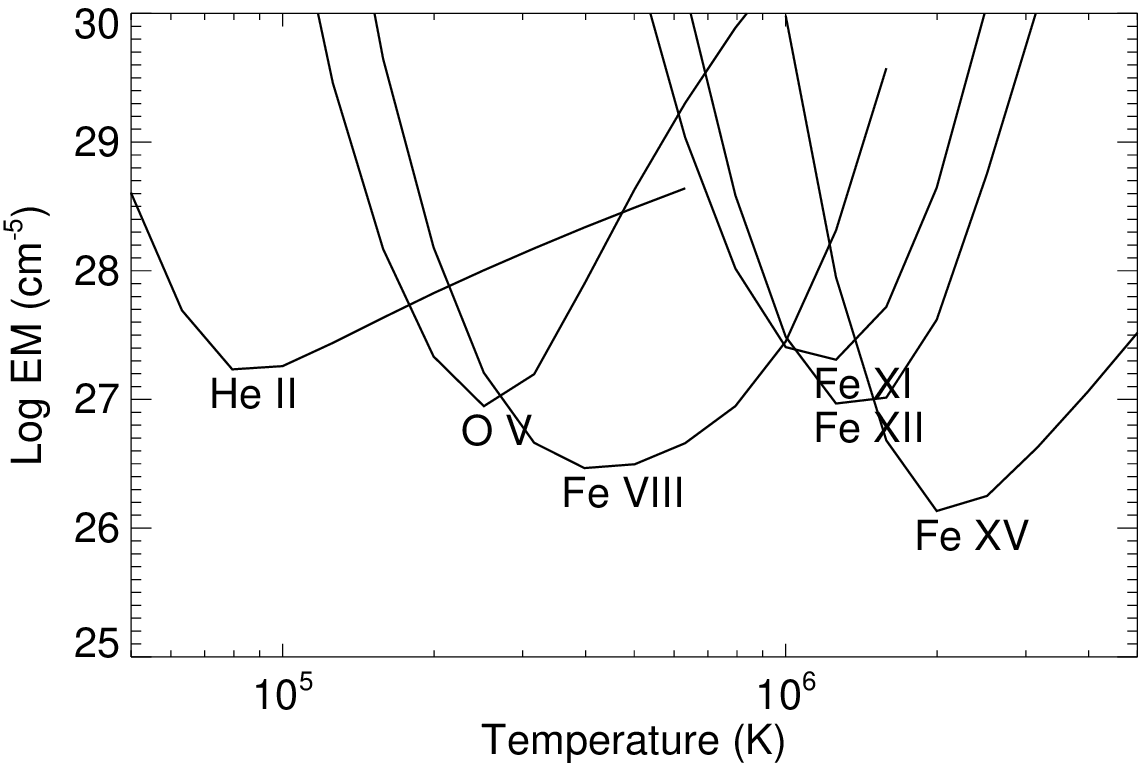}
 \caption{Loci plot derived from emissions within the microflare QR1.}
 \label{fig:loci01}
\end{figure}

\begin{figure}
 \centering
 \includegraphics[width=8cm]{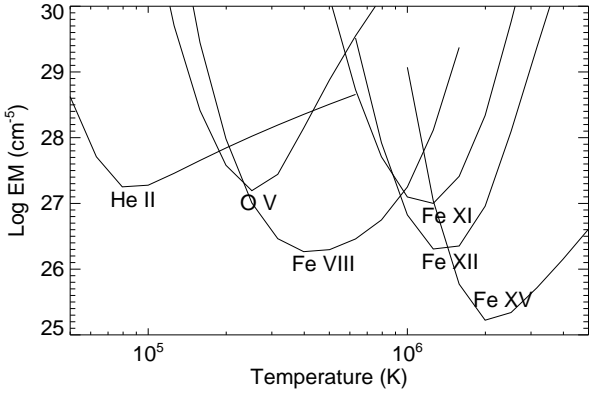}
 \caption{Loci plot derived from emissions within the microflare CH1.}
 \label{fig:loci08}
\end{figure}

The temperature distribution of the microflare is studied by
using emission lines over a wide temperature range.
A contribution function of each emission line is calculated
using the CHIANTI atomic database and applying coronal abundance
determined by \citet{feldman1992}.
The column emission measure (EM) is derived by the equation
\begin{equation}
EM = I / g(T,n_e)
\label{eq:em}
\end{equation}
where $I$ is the observed radiance.
and $g(T,n_e)$ denotes the contribution function of the emission line.
Figures \ref{fig:loci01} and \ref{fig:loci08} present
loci plots of the microflares.
Each curve presents a column emission measure assuming
that the measured emission originates from an isothermal plasma
at a given temperature.
If the emitting plasma is purely isothermal, all the curves should
meet at one temperature, which is not the case here.
Convergence of multiple curves around $1\times10^6$ K implies that
the emission measure peaks at that temperature.
\citet{warren2009} pointed out 
that the emission measure is strongly peaked at $1\times10^6$ K
in quiet regions, which is consistent with our results.
The \ion{Fe}{xv} curve reaches higher than
the curves of \ion{Fe}{viii}  \ion{Fe}{xi}, and \ion{Fe}{xii} at $1\times10^6$ K,
indicating an additional high temperature component around $2\times10^6$ K
in the microflare.
A sporadic emission detected
with XRT Be\_thin filters in Fig. \ref{fig:image}
also supports this scenario.
The fact that no emission was detected in
\ion{Fe}{xvii} suggests the highest temperature
is below $4\times10^6$ K.
Differential emission measure (DEM) analysis would give
us more detailed information,
however, a DEM analysis was not possible with our data set
because of the small number of emission lines available.

\subsection{Radiative loss rate}

Assuming an isothermal plasma, the radiative loss from
microflares is estimated by using a total radiative loss function.
Since the emission measure of the microflare peaks around $1\times10^6$ K
as discussed above, the column emission measure at that
temperature is derived from Equation \ref{eq:em}
by using the \ion{Fe}{xii} $\lambda$19.512~nm radiance at the
peak time of the lightcurves.
The column emission measure is 
multiplied by the apparent area of the microflare
to obtain a total emission measure.
A radiative loss function is derived from the CHIANTI atomic database
assuming a typical density of $3\times10^9$ cm$^{-3}$
from Table \ref{table:bp}.
The radiative loss function is estimated to be
$4\times10^{-22}$ erg cm$^3$ s$^{-1}$
at the temperature of $1\times10^6$ K.
Radiative loss rates of microflares are calculated by
multiplying the radiative loss function and the
total emission measure,
which are listed in the last column of Table \ref{table:bp}.
The radiative loss rates range from
$2\times10^{21}$ to $6\times10^{22}$ erg s$^{-1}$.
As shown in Figs. \ref{fig:loci01} and \ref{fig:loci08},
actual microflares are not isothermal.
Therefore, our results are considered as the
lower limit of the radiative loss rate
since the remaining component could also contribute to
the radiative loss.

\subsection{Estimated cooling times}

The observed delay times between the peaks in X-ray and cool emissions
are in the 8 -- 21 min range (Table \ref{table:bp}).
Theoretical cooling times are computed to determine if the delay times
can be attributed to the cooling of hot coronal loops.
\citet{cargill1995} derived an analytical solution for the cooling time
from a combination of conductive cooling and radiative cooling,
taking into account chromospheric evaporation.
It requires temperature $T_0$, density $n_e$, and loop length $L$
as initial parameters
(from Equation 14E in \citealt{cargill1995}).
\begin{equation}
\tau = 2.35\times 10^{-2} \frac{L^{5/6}}{T_0^{1/6} n_e^{1/6}}
\label{eq:cool}
\end{equation}
Although the internal structure of microflares is not resolved in our
observations, the loop size is assumed to be an arc of an apparent diameter
of the microflare.
The density is estimated from the \ion{Fe}{xii} emission ratio analysis
in the previous section.
The initial temperature is set to $2\times10^6$ K,
the temperature of maximum fractional ionization for \ion{Fe}{xv}.
The cooling time is calculated down to $10^5$ K.
The estimated cooling times are presented in Table \ref{table:bp},
which is 9 min for QR1 and 8 min for CH1.

Judging from the lightcurves, the observed time delay between the peaks in the
X-ray and cool emission lines is 12 min for QR1 and 15 min for
CH1 (Figs. \ref{fig:lc01} and \ref{fig:lc08}).
This discrepancy between estimated cooling time and observed delay times
is comparable to the crude temporal resolution of the rastering spectrometers,
which may have missed the real emission peak.
The uncertainty of the loop length estimation could cause error in
the cooling time calculation, since the loop was not resolved
in XRT images.
If the actual loop was longer than our estimation,
the cooling time could be longer as it largely depends on $L$
(Equation \ref{eq:cool}).
As an alternative, the coincidence of independent brightenings
in the transition region could be an explanation.
But considering the fact that the delayed cool emissions are observed in
7 out of 10 events (Fig \ref{fig:lc_all}), the cool emissions are likely
to be related to microflares, rather than coincidence of other
brightening events in transition region.
Another possibility is that the simplified loop cooling
model might be inadequate to reproduce the observations.

The estimated cooling times of microflares are much shorter than
those of large scale flares.
This is explained by the short conductive cooling time scale
of small coronal loops expressed as
$\tau_{cond} = 4\times10^{-10} n_e L^2 T_0^{-5/2}$ 
where $n_e$, $L$, and $T_0$ respectively denote density, loop length,
and temperature \citep{cargill1995}.
Shorter loop lengths result in shorter conductive cooling time,
while radiative cooling does not depend on the loop length.

\section{Discussion}

The temporal evolution of microflares is traced in quiet regions
and in polar coronal holes.
The lightcurves exhibit impulsive increases in emission over a wide temperature
ranging from \ion{He}{ii} to \ion{Fe}{xv}.
Hot emission peaks in X-ray, \ion{Fe}{xv}, and \ion{Fe}{xii} were
followed by enhancements in cool emission in \ion{He}{ii} or
in \ion{O}{v} for 7 out of 10 microflares.
The delay time between the peaks in X-ray and cool emission
was 8 -- 22 min.
Such delayed peaks in cool emission are
common to large flares \citep{curdt2004,teriaca2006,raftery2009},
although the time delays vary from tens of minutes to hours.
The lightcurve behavior is interpreted as the cooling of hot plasma
produced by impulsive heating at the beginning of the flare.
To test this hypothesis, observed time delays are compared with
the simple loop cooling model by \citet{cargill1995}.
Although the temporal resolution is not better than 6 min,
theoretical cooling times generally agree with observations.
Therefore, lightcurves of microflares can be
explained as impulsive heating followed by the cooling of
small coronal loops.
The small loop length of microflare compared to
that of active regions results in a shorter conductive cooling time,
while the radiative cooling time is independent of the loop length.
The short time scale of microflares is largely due to
the small length of the loops.

\citet{benz1999} investigated the cross-correlations between
lightcurves of different emission lines and claimed
that \ion{He}{I} and \ion{O}{v} peak emission precedes
the \ion{Fe}{xii} peak emission by 5 min, although the correlation
coefficients were low.
They infer that the chromosphere is heated to coronal
temperatures by small heating events.
We did not find noticeable peaks in cool emissions
prior to the coronal emission peak.
A possible reason is that the temporal resolution of our
observations was insufficient to resolve rapidly changing phenomena.
Nevertheless, enhanced cool emission was found
at the time of peak X-ray flux,
which could be a signature of chromospheric evaporation
analogous to large flares.
\citet{benz1999} also reported that the \ion{Fe}{ix/x} lightcurve 
lags the \ion{Fe}{xii} lightcurve by 23 s, which was interpreted as
coronal plasma cooling.

It is interesting to note that lightcurves of
microflares in quiet regions and in coronal holes have common
characteristics.
Delayed cool emission in a bright point associated with
a polar coronal jet was reported \citep{culhane2007a}.
In our study, the delayed emission was found
both in coronal holes and in quiet regions.
It suggests that the cooling of microflares in coronal hole and
in quiet region are explained by the same mechanism.

The line broadening of spectra recorded during
microflares can be interpreted as superposition of
unresolved flows of up to $-200$ km~s$^{-1}$.
\citet{mariska2007} found an impulsive intensity increase
accompanied by blueshifts in \ion{Fe}{xii} and \ion{Fe}{xv}
in an active region loop.
An impulsive heating could produce a flow along the loop.
The flow would be observed as a line broadening rather than
a distinct flow if coronal structures are smaller than
resolution of the instrument.
\citet{berkebile2009} detected upflows and downflows of
$10$ -- $40$ km~s$^{-1}$
in \ion{He}{i}, \ion{O}{v}, and \ion{Ne}{vi}
for different microflares near disk center.
But no distinct flows were found from our observations
in \ion{He}{ii} or \ion{O}{v}.
A possible reason is that the flows were almost vertical
in the transition region and thus not detected near the
limb in our observations.
The broadened spectra of \ion{Fe}{xv}
are interpreted as flows along the coronal loop
(Fig. \ref{fig:pro01}).

We found increases in density accompanying the increase in emission,
which supports chromospheric evaporation models.
Increased coronal densities have been found in previous studies
with EIS.
\citet{chifor2008} showed a density dependence on the upflow velocity
in a coronal jet.
\citet{watanabe2009} derived a high density of
$\geq 10^{11}$~cm$^{-3}$ in a flare loop from
\ion{Fe}{xiii} line ratio analysis.
Our study showed the temporal variation of the coronal density
associated with microflares.
It implies that impulsive heating led to the chromospheric
evaporation.
The evaporation flow along the coronal loop could contribute
to the line broadenings mentioned above.

Although the properties of microflares are quite similar
in quiet regions and in coronal holes,
a notable difference is found in their surroundings.
Microflares in quiet regions are accompanied by coronal dimmings,
while microflares in coronal holes are associated with jets.
This is probably due to a difference in the magnetic field configuration
of the surroundings.
Magnetic loops in bright points interact with
open fields in coronal holes to produce coronal jets
\citep{shibata1994}.
The hot plasma ejected along the open fields caused
a diffuse jet above the bright point in Fig. \ref{fig:image}.
\citet{pariat2009} performed 3D MHD simulations of
a twisted emerging flux rope reconnecting with open magnetic fields.
They claimed that the reconnection underwent two distinctive regimes:
an impulsive mode and a quasi-steady mode \citep{pariat2010}.
We speculate that the impulsive reconnection produces microflares,
while the quasi-steady reconnection maintained
persistent emission in the coronal bright point.
In our observations, no clear signature of a jet is seen in
the spectra. It is probably due to a faint emission from the jet and
short exposure time of our observations.
Although the jet is noticeable in difference images in Fig. \ref{fig:image},
a dominant emission comes from a background corona,
hence the spectrum did not indicate a significant
Doppler shift as a whole.

The microflares and associated coronal dimmings in quiet regions
are essentially the same as those observed in active regions
\citep{innes2009,innes2010}.
Quiet regions are supposedly filled with
closed magnetic fields \citep{ito2010}.
In the case of bright point QR1, the coronal dimming started in
the surroundings when the emission began to rise in the bright point.
The changes in magnetic field configuration which were initiated by reconnection
are likely to be the cause of coronal dimmings in the surroundings.
For large scale flares, it has been established that coronal
dimmings are associated with CMEs by statistical studies
\citep{bewsher2008, reinard2008}.
A possible explanation for the coronal dimming is
the removal of overlying coronal field \citep{innes2010}.
\citet{archontis2010b} used 3D MHD simulations to
study flux emergence into horizontal fields.
They demonstrated pre-existing magnetic fields in the corona
are peeled off via reconnections with the emerging flux.
We speculate that coronal magnetic fields are removed
through reconnections at microflares to form the coronal dimmings.
In our dataset no clear signature of eruption is detected.
A possible reason is that it is not easy to identify eruptions 
in \ion{He}{ii} 30.4 nm on-disk images,
because the background emission is highly structured and dynamic.
In addition, the 10 min cadence of our {\it STEREO} dataset
might not be sufficient to detect short-lived eruptions.
Microflares are often accompanied by cool eruptions.
\citet{kamio2010a} reported an X-ray jet associated with
a cool eruption in \ion{He}{ii} 30.4 nm above the limb.
Further observations with higher cadence are needed
to establish a relationship between microflares and cool eruptions.

We focused on microflares in quiet regions and coronal holes,
but microflares in active regions indicate higher temperatures.
In our study, the hottest emission line in microflares
is \ion{Fe}{xv} at $2\times10^6$ K.
\citet{shimizu1995} obtained a temperature range of
$4\times10^6\sim8\times10^6$ K
from a statistical study of microflares in active regions.
\citet{brosius2009} detected \ion{Fe}{xix} emission ($8\times10^6$ K) in
a microflare in an active region.
Scaling laws of the flare peak temperature and emission measure have been
determined \citep{feldman1996, shibata1999b}, implying that 
different scales of flares are produced by the same mechanism.
If the relationship could be extended to a lower emission measure,
small events are expected to show a low temperature.
\citet{aschwanden2008} derived a scaling law of $EM \propto T^{4.7}$
from microflares to large scale flares.
Although the number of events detected in our study
is limited, their scaling law
can be compared with our results at $1\times10^6$ K.
The scaling law by \citet{aschwanden2008} estimates
an emission measure of $10^{44}$~cm$^{-3}$ at $1\times10^6$ K,
which is close to $10^{42} \sim 10^{44}$~cm$^{-3}$ from our results.
Probably improved resolution and sensitivity of instruments used
allowed us to detected small events in quiet regions
and in coronal holes.
Although we selected bright points which were obvious in the X-ray images,
cooler events which would not show up in X-ray were also reported.
\citet{young2007b} reported an active region brightening
which was visible up to \ion{Fe}{xii}.
Therefore, transient brightenings with
a lower temperature than our study have been shown to exist.

\section{Conclusions}
We traced the evolution of microflares in coronal bright points.
The lightcurves of microflares show a common behavior
in quiet regions and in coronal holes,
starting with a impulsive rise in a wide temperature range
up to \ion{Fe}{xv} at $2\times10^6$~K.
The spectra at the impulsive peak exhibited broad profiles,
which are interpreted as a superposition of flows caused
by the microflares.
After the hot coronal emissions decreased to normal level,
enhanced emissions were noticed in cool emissions with
a delay of 8--22 min.
The observed delay times agree with
a simple cooling model of a coronal loop \citep{cargill1995},
though the temporal resolution of the measurement is not high enough.
The line ratio analysis indicated a density increase coinciding with
an impulsive emission peak, supporting chromospheric evaporation
models.

These characteristics of microflares are common to active region
flares, which suggests that microflares and large flares are
produced by the same mechanism.
The empirical flare scaling law \citep{aschwanden2008} falls in the 
range of the observed emission measure and temperature of 
microflares.
A notable difference is found in the surroundings of microflares;
diffuse coronal jets are produced above bright point in coronal holes
while coronal dimmings are formed in quiet regions.
They are thought to be due to differences in the magnetic field
configuration of the surroundings of the microflares.
An interaction of an emerging flux with open fields and closed
fields must be examined in detail by theoretical studies.

\begin{acknowledgements}
We would like to thank the referee for constructive suggestions.
{\it Hinode} is a Japanese mission developed and launched by ISAS/JAXA,
with NAOJ as domestic partner and NASA and STFC (UK) as international
partners.
It is operated by these agencies in co-operation with ESA and NSC (Norway).
{\it STEREO} is a project of NASA. The data from the SECCHI instrument
used here was produced by an international consortion led by the
Naval Research Lab (USA).
The STEREO contributions by the MPS were supported by DLR grant 50OC0501.
\end{acknowledgements}

\bibliographystyle{aa}
\bibliography{reference.bib}

\end{document}